\documentclass[]{mn2e}
\usepackage{epsfig}
\def\lsim{\vcenter{\hbox{$<$}\offinterlineskip\hbox{$\sim$}}}
\def\gsim{\vcenter{\hbox{$>$}\offinterlineskip\hbox{$\sim$}}}
\title[The AGB superwind speed at low metallicity]
{The AGB superwind speed at low metallicity}
\author[Jonathan R. Marshall et al.]{Jonathan R. Marshall$^{1}$\thanks{E-mail:
jrm@astro.keele.ac.uk}, Jacco Th. van Loon$^{1}$, Mikako Matsuura$^{2}$,
\newauthor
Peter R. Wood$^{3}$, Albert A. Zijlstra$^{2}$ and Patricia A.
Whitelock$^{4}$\\
$^{1}$Astrophysics Group, School of Chemistry and Physics,
Keele University, Staffordshire, ST5 5BG, United Kingdom\\
$^{2}$UMIST, Department of Physics, P.O. Box 88, Manchester,
M60 1QD, United Kingdom\\
$^{3}$Research School of Astronomy and Astrophysics, Australian National
University, Cotter Road, Weston Creek, ACT 2611, Australia\\
$^{4}$South African Astronomical Observatory, P.O. Box 9, 7935 Observatory,
      South Africa}
\date{Submitted 2004}
\pagerange{\pageref{firstpage}--\pageref{lastpage}}
\pubyear{2004}
\begin{document}
\maketitle
\label{firstpage}
\begin{abstract}
We present the results of a survey for OH maser emission at 1612 MHz from
dust-enshrouded AGB stars and supergiants in the LMC and SMC, with the Parkes
radio telescope, aimed at deriving the speed of the superwind from the
double-peaked OH maser profiles. Out of 8 targets in the LMC we detected 5, of
which 3 are new detections --- no maser emission was detected in the two SMC
targets. We detected for the first time the redshifted components of the OH
maser profile in the extreme red supergiant IRAS\,04553$-$6825, confirming the
suspicion that its wind speed had been severely underestimated. Despite a much
improved spectrum for IRAS\,04407$-$7000, which was known to exhibit a
single-peaked OH maser, no complementary peak could be detected. The new
detection in IRAS\,05003$-$6712 was also single-peaked, but for two other new
detections, IRAS\,04498$-$6842 and IRAS\,05558$-$7000, wind speeds could be
determined from their double-peaked maser profiles. The complete sample of
known OH/IR stars in the LMC is compared with a sample of OH/IR stars in the
galactic centre. The LMC sources generally show a pronounced asymmetry between
the bright blueshifted maser emission and weaker redshifted emission, which we
attribute to the greater contribution of amplification of radiation coming
directly from the star itself as the LMC sources are both more luminous and
less dusty than their galactic centre counterparts. We confirm that the OH
maser strength is a measure of the dust (rather than gas) mass-loss rate. At a
given luminosity or pulsation period, the wind speed in LMC sources is lower
than in galactic centre sources, and the observed trends confirm simple
radiation-driven wind theory if the dust-to-gas ratio is approximately
proportional to the metallicity.
\end{abstract}

%=========================================================================== 1
\section{Introduction}

Asymptotic Giant Branch (AGB) stars lose $\sim$35 to 85\% of their mass before
ending up as white dwarfs. Much of this mass loss takes place near the tip of
the AGB during the superwind stage, at mass-loss rates of up to
$\dot{M}\sim10^{-4}$ M$_\odot$ yr$^{-1}$ (van Loon et al.\ 1999, 2003) leading
to the early cessation of their stellar evolution and resulting in chemical
enrichment of the interstellar medium (ISM). The mass loss occurs in the form
of a stellar wind reaching typical outflow velocities of 5 to 30 km s$^{-1}$.
These winds are driven by radiation pressure on dust grains that form at a
height of several stellar radii (Goldreich \& Scoville 1976) in the presence
of strong radial pulsation of the base of the stellar atmosphere. Likewise,
more massive stars ($M_{\rm ZAMS}\gsim8$ M$_\odot$) may pass through a red
supergiant phase and lose mass in a similar manner.

%
% TABLE 1
%
\begin{table*}
\caption[]{The target sample plus the detections from Wood et al.\ (1992).
Bolometric magnitudes $M_{\rm bol}$, pulsation periods $P$, K-band pulsation
amplitudes ${\Delta}K$, K-band magnitudes and colours $J-K$ are from Whitelock
et al.\ (2003), {\it IRAS} 25 $\mu$m flux densities $F_{25}$ (not
colour-corrected) are from Trams et al.\ (1999), and spectral types are from
van Loon et al.\ (in preparation), except where noted otherwise. We adopt
distances to the SMC and LMC of 50 and 60 kpc, respectively (distance moduli
of 18.5 and 18.9 mag, respectively). Also listed are the previously detected
OH masers, and whether we detected OH maser emission here (no entry means
never tried).}
\begin{tabular}{llcccccclrr}
\hline\hline
Object           &
Alternative      &
$M_{\rm bol}$    &
$P$              &
${\Delta}K$      &
$K$              &
$F_{25}$         &
$J-K$            &
Spectral         &
Previous         &
This work        \\
name             &
name             &
(mag)            &
(mag)            &
(days)           &
(mag)            &
(Jy)             &
                 &
type             &
OH               &
OH               \\
\hline
{\it SMC}\\
IRAS\,00483$-$7347 &
LI-SMC\,61         &
$-$7.20\rlap{$^8$} &
1200\rlap{$^8$}    &
0.61\rlap{$^8$}    &
9.10\rlap{$^3$}    &
0.46\rlap{$^3$}    &
3.01\rlap{$^3$}    &
late-M             &
                   &
no                 \\
IRAS\,00591$-$7307 &
HV\,11417          &
$-$8.30\rlap{$^2$} &
\llap{$\gsim$}1000\rlap{$^1$} &
\llap{$>$}2.00     &
8.30\rlap{$^3$}    &
0.10\rlap{$^3$}    &
1.33\rlap{$^3$}    &
M5e I\rlap{$^1$}   &
                   &
no                 \\
\hline
{\it LMC}\\
IRAS\,04407$-$7000 &
LI-LMC\,4          &
$-$7.11            &
1199               &
1.23               &
8.79               &
0.76               &
2.34               &
M7.5               &
yes\rlap{$^6$}     &
yes                \\
IRAS\,04498$-$6842 &
LI-LMC\,60         &
$-$7.72            &
1292               &
1.30               &
8.08               &
0.89               &
1.86               &
M10                &
                   &
yes                \\
IRAS\,04509$-$6922 &
LI-LMC\,77         &
$-$7.28            &
1292               &
1.45               &
8.59               &
0.86               &
2.21               &
M10\rlap{$^6$}     &
no\rlap{$^9$}      &
no                 \\
IRAS\,04516$-$6902 &
LI-LMC\,92         &
$-$7.11            &
1091               &
1.41               &
8.72               &
0.55               &
2.32               &
M9\rlap{$^6$}      &
                   &
no                 \\
IRAS\,04553$-$6825 &
WOH\,G064          &
$-$9.19\rlap{$^5$} &
\hspace*{1mm}841   &
0.34               &
7.09               &
\llap{1}3.53       &
2.60\rlap{$^4$}    &
M7.5               &
yes\rlap{$^9$}     &
yes                \\
IRAS\,05003$-$6712 &
LI-LMC\,0297       &
$-$6.20            &
\hspace*{1mm}883   &
1.59               &
9.95               &
0.33               &
2.95               &
M9                 &
                   &
yes                \\
IRAS\,05294$-$7104 &
LI-LMC\,1153       &
$-$6.79            &
1079               &
1.20               &
9.21               &
0.56               &
2.97               &
M8                 &
                   &
no                 \\
IRAS\,05558$-$7000 &
LI-LMC\,1790       &
$-$6.97            &
1220               &
1.42               &
9.25               &
0.80               &
3.27               &
                   &
                   &
yes                \\
\hline
\multicolumn{5}{l}{\it Sample from Wood et al.\ (1992) --- excluding
IRAS\,04553$-$6825} \\
IRAS\,04545$-$7000 &
LI-LMC\,159        &
$-$6.56            &
1216               &
1.57               &
\llap{1}0.13       &
0.83               &
5.70\rlap{$^9$}    &
                   &
yes\rlap{$^9$}     &
                   \\
IRAS\,05280$-$6910 &
NGC\,1984-IRS1     &
$-$7.75\rlap{$^9$} &
                   &
                   &
8.19\rlap{$^9$}    &
\llap{2}4.18\rlap{$^7$} &
                   &
                   &
yes\rlap{$^9$}     &
                   \\
IRAS\,05298$-$6957 &
LI-LMC\,1164       &
$-$6.72\rlap{$^9$} &
1280\rlap{$^9$}    &
2.00\rlap{$^9$}    &
\llap{1}0.29\rlap{$^9$} &
1.38               &
3.54\rlap{$^9$}    &
                   &
yes\rlap{$^9$}     &
                   \\
IRAS\,05329$-$6708 &
LI-LMC\,1286       &
$-$6.95            &
1262               &
1.50               &
9.90               &
1.23               &
5.10\rlap{$^9$}    &
                   &
yes\rlap{$^9$}     &
                   \\
IRAS\,05402$-$6956 &
                   &
$-$6.77            &
1393               &
1.80               &
\llap{1}0.40       &
1.02               &
4.46\rlap{$^9$}    &
                   &
yes\rlap{$^9$}     &
                   \\
\hline
\end{tabular}\\
References: $^1$ Elias, Frogel \& Humphreys (1980), $^2$ Elias, Frogel \&
Humphreys (1985), $^3$ Groenewegen \& Blommaert (1998), $^4$ Trams et al.\
(1999), $^5$ van Loon et al.\ (1999), $^6$ van Loon et al.\ (1998a), $^7$ van
Loon et al.\ (2001b), $^8$ Whitelock et al.\ (1989), $^9$ Wood et al.\ (1992).
\end{table*}

During the superwind stage the circumstellar dust envelope becomes opaque at
visual wavelengths, but the absorbed stellar light is re-radiated at infrared
(IR) wavelengths. The resulting spectral energy distribution may be modelled
in order to derive the mass-loss rate, provided we know the dust-to-gas ratio,
dust grain properties, stellar luminosity and effective temperature, and
outflow kinematics. Data on Magellanic Cloud and galactic dust-enshrouded AGB
stars are consistent with a dust-to-gas ratio which depends linearly on
metallicity, without a strong metallicity dependence of the total (gas+dust)
mass-loss rate (van Loon 2000). This relies on the validity of simple
dust-driven wind theory, in which the wind speed, $v_{\rm exp}$, depends on
the dust-to-gas ratio, $\psi$, and luminosity, $L$, as (van Loon 2000; cf.\
Habing, Tignon \& Tielens 1994; Elitzur \& Ivezi\'{c} 2001):
\begin{equation}
v_{\rm exp}\propto\psi^{1/2}L^{1/4}.
\end{equation}
It is thus of importance, both for deriving accurate mass-loss rates as well
as for understanding the mass-loss mechanism, to measure wind speeds for stars
of different luminosity and metallicity.

The wind speed may be measured from molecular emission line profiles at radio
and (sub)mm wavelengths. This is best done for the abundant CO molecule, but
its lines are too weak to detect outside the local galactic disk with
presently available observatories. Oxygen-rich dust-enshrouded AGB stars
(invariably of M spectral type) may also exhibit intense maser emission from
the stimulated re-emission of stellar and circumstellar radiation by SiO,
H$_2$O and/or OH molecules. These masers are radially stratified due to the
spatial separation of the molecules in the outflowing circumstellar envelope
(see Habing (1996) for a review). SiO is easily condensed into dust grains and
therefore is abundant only in the dust-free inner cavity close to the star.
Its maser emission is tangentially amplified resulting in a single peak which
can be used to determine the stellar velocity. H$_2$O masers originate in the
dust formation zone. In Miras these are predominantly amplified tangentially,
providing another means of measuring the stellar velocity, but in the more
extreme OH/IR stars also radial amplification is effective, resulting in a
double-peaked line profile from which the outflow velocity can be measured
(Takaba et al.\ 1994). OH is formed by the dissociation of H$_2$O by
interstellar UV radiation (Goldreich \& Scoville 1976). As the OH maser
emission is radially amplified it provides a means for measuring the speed of
the fully developed wind (Elitzur, Goldreich \& Scoville 1976) --- as opposed
to the H$_2$O masers which probe a region where the wind has not yet reached
its final speed (Richards, Yates \& Cohen 1998, 1999).

OH masers have been widely detected in galactic metal-rich samples. The best
opportunities to investigate the wind speed in a low-metallicity environment
are provided by the Large Magellanic Cloud (LMC) and Small Magellanic Cloud
(SMC), with metallicities of the intermediate-age stellar populations of
$\sim\frac{1}{3}$ and $\sim\frac{1}{5}$ solar respectively. The Magellanic
Clouds have the additional advantage that their distances are reasonably well
known, and hence the dependence of the wind speed on stellar luminosity can be
determined. Prior to this study, only seven identified OH/IR stars were known
in the LMC, of which five are AGB stars (the other two are supergiants), and
none in the SMC. Amongst the five AGB stars, one was detected as a
single-peaked OH maser line, and only one showed an uncomplicated
``classical'' double-peaked profile yielding a reliable value for its wind
speed. We here present new detections of magellanic OH/IR stars, and an
analysis of the dependence of the superwind speed on stellar luminosity and
metallicity.

%
% TABLE 2
%
\begin{table*}
\caption[]{OH maser detections, with the heliocentric velocities of the blue
and red peaks (where identified), corresponding wind speed $v_{\rm exp}$ from
the separation of the peaks, peak flux density and integrated flux $F_{\rm
int}$. The rms noise levels in the flux densities and in the integrated fluxes
are also listed as $\sigma$ and $\sigma_{\rm int}$, respectively (for
IRAS\,04407$-$7000, IRAS\,04498$-$6842, IRAS\,05003$-$6712 and
IRAS\,05558$-$7000 these values refer to the smoothed spectra shown in Figs.\
1--4, with an effective channel width of 0.36 km s$^{-1}$).}
\begin{tabular}{lclrclrccc}
\hline\hline
Object                                &
Integration                           &
\multicolumn{2}{c}{Velocity at peaks} &
$v_{\rm exp}$                         &
\multicolumn{2}{c}{Peak flux density} &
$\sigma$                              &
$F_{\rm int}$                         &
$\sigma_{\rm int}$                    \\
name                                  &
time (s)                              &
\multicolumn{2}{c}{(km s$^{-1}$)}     &
(km s$^{-1}$)                         &
\multicolumn{2}{c}{(mJy)}             &
(mJy)                                 &
(mJy km s$^{-1}$)                     &
(mJy km s$^{-1}$)                     \\
                                      &
                                      &
Blue                                  &
Red                                   &
                                      &
Blue                                  &
Red                                   &
                                      &
                                      &
                                      \\
\hline
{\it LMC}\\
IRAS\,04407$-$7000 &
33000              &
240                &
270\rlap{?}        &
\llap{1}5\rlap{?}  &
 50                &
  8\rlap{?}        &
  5\rlap{.5}       &
\llap{1}15         &
10                 \\
IRAS\,04498$-$6842 &
28800              &
246                &
272                &
\llap{1}3          &
 23                &
 11                &
  5\rlap{.9}       &
34                 &
12                 \\
IRAS\,04553$-$6825 &
14455              &
252, 262           &
300, 280           &
\llap{2}4          &
\llap{6}00, 180    &
 70,  50           &
\llap{1}0\rlap{.5} &
\llap{2,9}50       &
13                 \\
IRAS\,05003$-$6712 &
32010              &
268                &
                   &
                   &
 33                &
$\sim$13\rlap{?}   &
  5\rlap{.1}       &
38                 &
12                 \\
IRAS\,05558$-$7000 &
32500              &
265                &
285                &
\llap{1}0          &
 17                &
 17                &
  6\rlap{.9}       &
50                 &
14                 \\
\hline
\multicolumn{6}{l}{\it Sample from Wood et al.\ (1992) --- excluding
IRAS\,04553$-$6825} \\
IRAS\,04545$-$7000 &
                   &
258                &
274                &
8                  &
\llap{1}40         &
 30                &
\llap{1}7          &
                   &
                   \\
IRAS\,05280$-$6910 &
                   &
255                &
289                &
\llap{1}7          &
 90                &
 70                &
\llap{1}7          &
                   &
                   \\
IRAS\,05298$-$6957 &
                   &
271                &
292                &
\llap{1}0\rlap{.5} &
\llap{2}40         &
130                &
\llap{1}7          &
                   &
                   \\
IRAS\,05329$-$6708 &
                   &
301                &
323                &
\llap{1}1          &
 70                &
130                &
\llap{1}7          &
                   &
                   \\
IRAS\,05402$-$6956 &
                   &
263                &
284                &
\llap{1}0\rlap{.5} &
 80                &
 60                &
\llap{1}7          &
                   &
                   \\
\hline
\end{tabular}
\end{table*}

%=========================================================================== 2
\section{Description of the sample}

The first maser searches of magellanic dust-enshrouded AGB stars and red
supergiants (Wood et al.\ 1992; van Loon et al.\ 2001b) were based on {\it
IRAS}-selected point sources (e.g.\ Loup et al.\ 1997) at a time when not much
was known about the nature of these {\it IRAS} sources. Since the mid-1990s,
in preparation for and based on {\it ISO} observations a considerable sample
of these sources have now been classified according to their luminosity,
variability, IR colours and chemical type (Zijlstra et al.\ 1996; van Loon et
al.\ 1998a; Groenewegen \& Blommaert 1998; Trams et al.\ 1999). As a result,
we are able to select detectable OH/IR candidates with greater certainty.
Their properties and those of previously detected magellanic OH/IR stars are
listed in Table 1. We adopt a distance to the LMC of $D_{\rm LMC}=50$ kpc, and
a distance to the SMC of $D_{\rm SMC}=60$ kpc.

The known OH maser sources all have extremely long pulsation periods
($P\gsim1200$ d) and large pulsation amplitudes (${\Delta}K\gsim1.5$ mag) ---
except the very luminous red supergiant IRAS\,04553$-$6825. They are all
enshrouded in a thick dust envelope, making them red at short wavelengths
($J-K\gsim3$ mag) and bright in the mid-IR ($F_{25}\gsim1$ Jy). It has
therefore been difficult to determine the optical spectral type for the OH/IR
stars, but where it has been possible they were of late-M type. Their
bolometric luminosities place them near or above the tip of the AGB ($M_{\rm
bol}\simeq-7$ mag). Hence targets were selected on the basis of being
oxygen-rich and luminous, having red colours and bright mid-IR flux densities,
and pulsating with very long periods and large amplitudes. In particular the
knowledge of the chemical type has allowed us to avoid carbon stars --- which
do not exhibit OH maser emission (many of the Wood et al.\ (1992) targets have
later been found to be carbon stars). Unfortunately, not many such OH/IR
candidates are known in the SMC; presumably because the potential candidates
turn into carbon stars in this metal-deficient environment.

%=========================================================================== 3
\section{Observations}

The 64m radio telescope at Parkes, Australia, was used from 14 to 22 August,
2003, to observe the OH satellite line at 1612 MHz. The multibeam receiver was
used in two orthogonal polarizations, each of 8 MHz bandwidth and sampled with
8192 channels yielding 0.18 km s$^{-1}$ channel$^{-1}$.

Most of the observations were performed in frequency switching mode with a
frequency throw of 0.5 MHz (93 km s$^{-1}$ at 1612 MHz), with integration
times of 5 sec in between. Total integration times amounted to about 15,000
sec, and double that for objects that were re-observed to confirm a suspected
detection. The first experiments on IRAS\,04498$-$6842, IRAS\,04509$-$6922 and
IRAS\,04553$-$6825 were performed in position switching mode, alternatingly
observing the source and patches of (presumed empty) sky at $30^\prime$ to the
North and South. Most of the integration time on these sources was however
obtained in frequency switching mode. IRAS\,04553$-$6825 was observed daily
and also used for flux calibration checks.

The data were processed into blocks of 250 sec worth of raw data with the
spectral line reduction programme {\sc spc}. Off-line reduction consisted of
combining the spectra with the various polarizations and frequency settings,
and baseline subtraction. A Fortran code was written to do this, but also to
first clean the spectra from human-made interference which can generate strong
spikes in the spectrum. These generally decay within a couple of integrations,
but they may persist in the average spectrum and be confused as a genuine
feature. There was a noticeable increase in the level of this ``spikey'' noise
during daytime.

The code, {\sc nospike}, works as follows: first a baseline is subtracted,
using the mode of 50 channels at a time. Then the mode of the squared values
is calculated to obtain a measure for the spread, $\sigma$, specifying a
minimum value for $\sigma$ to make it robust against anomalously low values in
the presence of strong signal. The programme then runs through every data
point and if a value is greater than a specified number of $\sigma$, that
value is replaced with the preceding value. The same code then averages the
polarizations and frequencies as appropriate (depending on the observing
mode), and stacks the observations, weighted according to exposure time and
the inverse square of system temperature (which was $\sim37$ K most of the
time).

The method of combining the spectra that provided the best signal-to-noise
ratio was the Total Power method, where the frequency-switched spectra were
aligned and averaged rather than ratio-ed first. This therefore did not remove
the baseline, which was done by subtracting a heavily smoothed version of the
original spectrum. This method was not robust enough for IRAS\,04553$-$6825
due to its extremely strong emission, and for this source the baseline was
instead fitted with a low-order polynomial function. The spectra of the OH
maser detections --- except IRAS\,04553$-$6825 --- were smoothed by a factor
two in order to reduce noise but preserve maser emission features.

%=========================================================================== 4
\section{Results}

%
% TABLE 3
%
\begin{table}
\caption[]{Non-detections, with values for the rms noise level in the original
spectrum and that which is obtained for any emission integrated over a 10 km
s$^{-1}$ interval.}
\begin{tabular}{lccc}
\hline\hline
Object             &
Integration        &
$\sigma$           &
$\sigma_{\rm int}$ \\
name               &
time (s)           &
(mJy)              &
(mJy km s$^{-1}$)  \\
\hline
{\it SMC}\\
IRAS\,00483$-$7347 &
13000              &
 9                 &
27                 \\
IRAS\,00591$-$7307 &
17000              &
 8                 &
24                 \\
\hline
{\it LMC}\\
IRAS\,04509$-$6922 &
24660              &
\llap{1}7          &
63                 \\
IRAS\,04516$-$6902 &
15500              &
\llap{1}1          &
47                 \\
IRAS\,05294$-$7104 &
16000              &
\llap{1}3          &
42                 \\
\hline
\end{tabular}
\end{table}

Tables 2 \& 3 show the obtained results for the OH maser detections and
non-detections, respectively. The wind speed $v_{\rm exp}$ is derived from the
separation of the blue and red peaks\footnote{We refer to the peak at the
lowest heliocentric velocity as the ``blue'' peak and the higher velocity
component as the ``red'' peak.}, ${\Delta}v_{\rm blue-red}$, rather than from
the separation of the extreme velocities where emission is detected (and which
is very hard to determine in noisy spectra). The integrated flux is obtained
by integrating over either the blue peak or the whole maser profile where the
red component is clearly identified. Upper limits for the emission from the
targets that failed to yield a clear maser detection may be taken to be
$3\sigma$. The individual detections and the non-detections are described in
separate subsections below.

The observed velocities of the bulk of the ISM seen in H~{\sc i} may in some
cases support or challenge claims for detected maser emission, as the OH/IR
stars are generally massive (hence young) enough to follow the kinematics of
the ambient ISM from which they have formed. Van Loon et al.\ (2001b)
demonstrated this to be the case for circumstellar masers in the LMC. Where we
refer to the H~{\sc i} velocities we have inspected the Parkes 21 cm Multibeam
Project\footnote{http://www.atnf.csiro.au/research/multibeam/} datacube along
the lines-of-sight towards our targets. For a global overview of the structure
and kinematics of the H~{\sc i} gas in the magellanic clouds the reader is
referred to Kim et al.\ (1998; LMC) and Stanimirovi\'{c} et al.\ (1999; SMC).

%========================================================================= 4.1
\subsection{IRAS\,04407$-$7000}

%
% FIGURE 1
%
\begin{figure}
\centerline{\psfig{figure=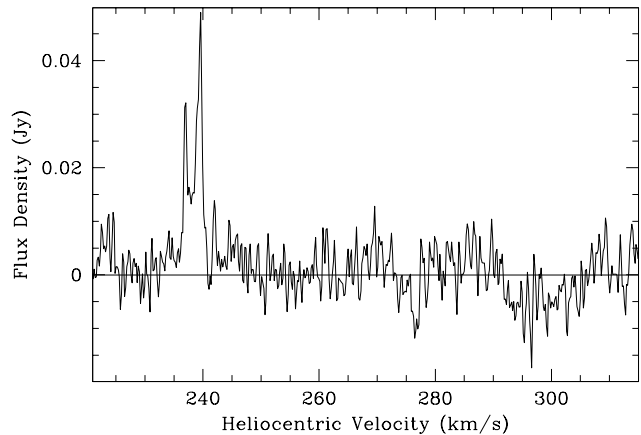,width=84mm}}
\caption[]{OH 1612 MHz maser emission from IRAS\,04407$-$7000.}
\end{figure}

IRAS\,04407$-$7000 is in all respects a typical massive dust-enshrouded
tip-AGB star. An integration with the AT Compact Array at high spatial
resolution led to the detection of a single-peaked 1612 MHz OH maser (van Loon
et al.\ 1998a) at a heliocentric velocity of 239 km s$^{-1}$. The peak flux
density was $F_{\rm peak}\sim50$ mJy, at 0.7 km s$^{-1}$ channel$^{-1}$ and an
rms noise level of $\sigma\sim12$ mJy.

Our much deeper integration with Parkes confirms the presence of strong OH
maser emission at 240 km s$^{-1}$, again peaking at $F_{\rm peak}\sim50$ mJy,
but also shows a secondary peak blueshifted by a few km s$^{-1}$ (Fig.\ 1).
This is probably substructure within one of the maser components --- most
likely the blue component (see Sect.\ 5). In one of the two observations for
this source a complementary peak at 270 km s$^{-1}$ was visible, which would
result in a wind speed of 15 km s$^{-1}$ (but we omit this estimate from
further analysis). The H~{\sc i} data along this line-of-sight was heavily
affected by emission in the sidelobes of the dish, which occurred at
velocities around 240 to 260 km s$^{-1}$. If this is also representative for
the neutral hydrogen near IRAS\,04407$-$7000, then these velocities are
consistent with our interpretation of the strong peak at 240 km s$^{-1}$ being
the blue component.

%========================================================================= 4.2
\subsection{IRAS\,04498$-$6842}

%
% FIGURE 2
%
\begin{figure}
\centerline{\psfig{figure=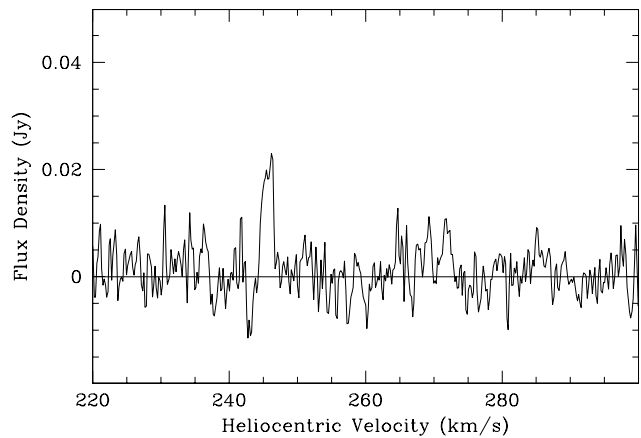,width=84mm}}
\caption[]{OH 1612 MHz maser emission from IRAS\,04498$-$6842.}
\end{figure}

IRAS\,04498$-$6842 is very luminous, but its long pulsation period, large
pulsation amplitude and fairly moderate mid-IR luminosity suggests it is an
AGB star rather than a supergiant. The high luminosity might result from Hot
Bottom Burning --- which also prevents AGB stars in the LMC that are more
massive than $\sim$4 M$_\odot$ from becoming carbon stars (van Loon et al.\
2001a). Or the bolometric luminosity calculated by Whitelock et al.\ (2003),
which was based on the limited {\it IRAS} data, may be an overestimate of the
true mean luminosity (indeed the luminosity determined from {\it ISO}
observations is lower).

Its OH spectrum (Fig.\ 2) shows a rather blunt blue-shifted peak at 246 km
s$^{-1}$, and a fainter red component at 272 km s$^{-1}$, which implies a wind
velocity of 13 km s$^{-1}$. Although the red component is at the limit of the
detection sensitivity, it was visible on both days on which IRAS\,04498$-$6842
was observed. The H~{\sc i} emission peaks sharply at $\sim$260 km s$^{-1}$,
which agrees very well with the velocities of the blue and red OH maser
components in IRAS\,04498$-$6842.

%========================================================================= 4.3
\subsection{IRAS\,05003$-$6712}

%
% FIGURE 3
%
\begin{figure}
\centerline{\psfig{figure=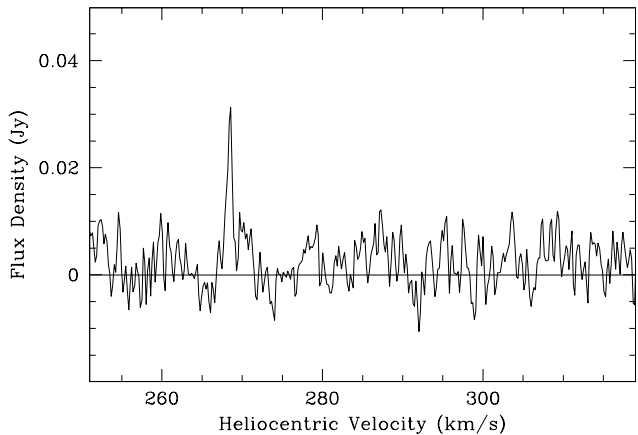,width=84mm}}
\caption[]{OH 1612 MHz maser emission from IRAS\,05003$-$6712.}
\end{figure}

Although IRAS\,05003$-$6712 is the faintest (bolometrically and at mid-IR
wavelengths) object and has one of the shortest pulsation periods amongst the
targets, it is still an extremely dust-enshrouded ($J-K=3$ mag; $F_{25}=0.33$
Jy), luminous ($M_{\rm bol}=-6.2$ mag) and cool (spectral type M9) AGB star
with strong (${\Delta}K=1.6$ mag) stellar pulsations. It is therefore not
unexpected that we detected OH maser emission from this source (Fig.\ 3).

The OH spectrum (Fig.\ 3) shows a sharp blue-asymmetric\footnote{By
``blue-asymmetric'' we mean that the flux density at the short wavelength side
of the centre of the line profile is higher than at the long wavelength side.}
peak at 268 km s$^{-1}$, but it is not clear where would be the corresponding
red component. The H~{\sc i} spectrum is affected by sidelobe interference
around 260 to 280 km s$^{-1}$, leaving emission from the main beam visible
around 280 to 300 km s$^{-1}$. This suggests indeed that the 268 km s$^{-1}$
OH maser peak is the blue component.

%========================================================================= 4.4
\subsection{IRAS\,05558$-$7000}

%
% FIGURE 4
%
\begin{figure}
\centerline{\psfig{figure=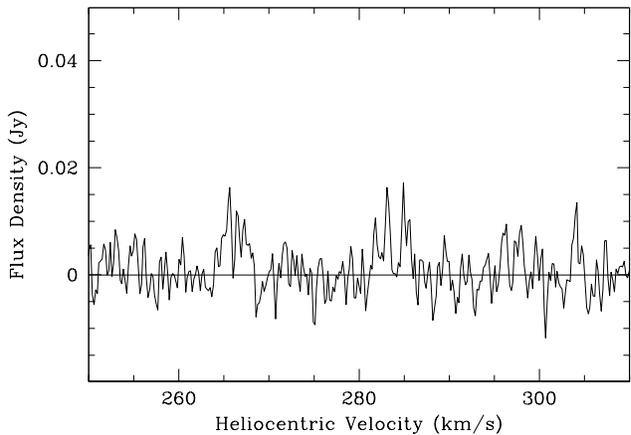,width=84mm}}
\caption[]{OH 1612 MHz maser emission from IRAS\,05558$-$7000.}
\end{figure}

IRAS\,05558$-$7000 is a typical dust-enshrouded tip-AGB star. With $J-K>3$ mag
it is also the reddest target, and its optical spectrum was too dim to
classify (note however that at least four of the Wood et al.\ (1992) OH/IR
stars are redder than IRAS\,05558$-$7000).

The OH spectrum of this source (Fig.\ 4) is unusual within this sample as both
the blue (at 265 km s$^{-1}$) and red (at 285 km s$^{-1}$) components are
equally bright (17 mJy). Situated in the outskirts of the LMC, the H~{\sc i}
emission (affected by sidelobe interference around 260 to 280 km s$^{-1}$)
extends to at least 350 km s$^{-1}$, and is therefore of limited help in
affirming the interpretation of the OH spectrum of IRAS\,05558$-$7000.

%========================================================================= 4.5
\subsection{The red supergiant IRAS\,04553$-$6825}

%
% FIGURE 5
%
\begin{figure}
\centerline{\psfig{figure=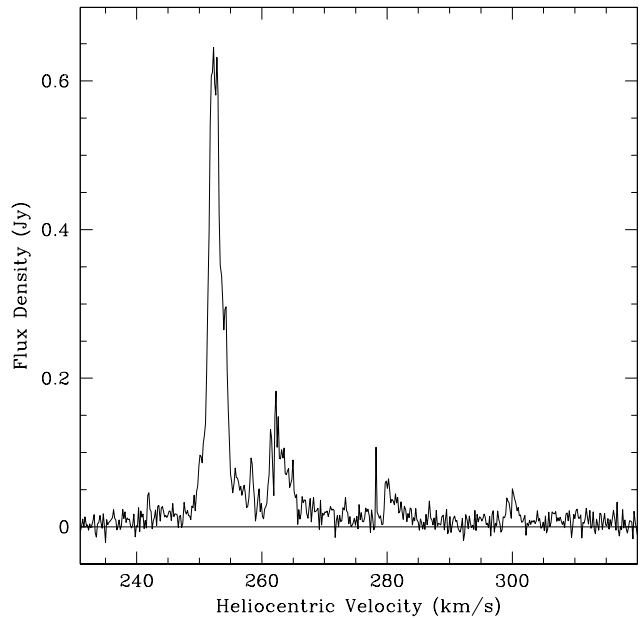,width=84mm}}
\caption[]{OH 1612 MHz maser emission from IRAS\,04553$-$6825.}
\end{figure}

The coolest, most luminous known LMC red supergiant, IRAS\,04553$-$6825 has
been much studied and was part of the sample observed by Wood et al.\ (1992).
It has been found to have extremely strong maser emission from several
sources: OH (Wood, Bessel \& Whiteoak 1986), H$_2$O (van Loon et al.\ 1998b)
and SiO (van Loon et al.\ 1996).

Detailed work on the H$_2$O and SiO maser emission by van Loon et al.\ (2001b)
indicated that the location of the star was at a greater velocity than
suggested by Wood et al.\ (1992), whose observations it turned out had shown
double-peaked structure within the blue-shifted component only. Our new data
clearly reveals two components, at 280 and 300 km s$^{-1}$ (Fig.\ 5), which
are red-shifted with respect to the previously detected OH maser emission as
well as red-shifted with respect to the peaks in the SiO and H$_2$O maser
emission which indicate the systemic (stellar) velocity at $v_\star\sim278$ km
s$^{-1}$ (Fig.\ 6) --- intriguingly, an extremely sharp OH emission spike was
detected at exactly this systemic velocity. The systemic velocity is confirmed
by the H~{\sc i} emission which peaks around $\sim$270 km s$^{-1}$.

A possible explanation for the multiple maser emission components would be the
presence of two expanding dust shells. The faster of these, with a wind speed
of $v_{\rm exp}\sim24$ km s$^{-1}$, is kinematically quite symmetrical with
respect to the systemic velocity. Similar OH maser profiles in other sources
have also been interpreted as due to bipolar outflows (Chapman 1988). Over the
two decades that have passed since the first spectrum, the blue-most peak has
increased in strength whilst the secondary blue component (at 262 km s$^{-1}$)
has diminished in strength. This may reflect changes in the structure of the
circumstellar matter distribution, but as maser emission is a highly
non-linear process it is very sensitive to small changes in the conditions
required for the population inversion and velocity coherency to occur.

%
% FIGURE 6
%
\begin{figure}
\centerline{\psfig{figure=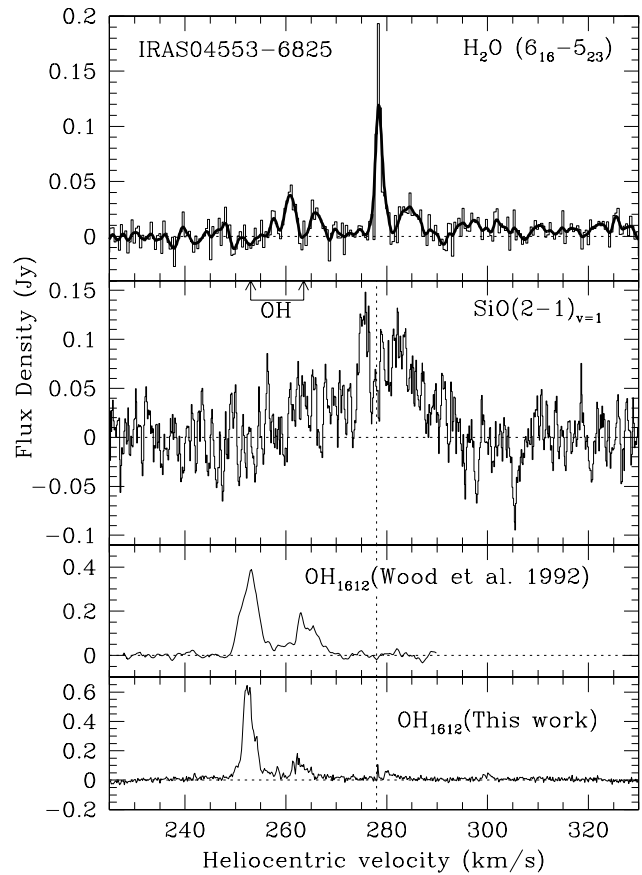,width=84mm}}
\caption[]{Comparison of the masers detected in IRAS\,04553$-$6825 (after van
Loon et al.\ 2001b).}
\end{figure}

%========================================================================= 4.6
\subsection{Non-detections}

On the basis of the (circum)stellar parameters the remaining three targets
IRAS\,04509$-$6922, IRAS\,04516$-$6902 and IRAS\,05294$-$7104 could have been
expected to exhibit OH maser emission at a similar level as the successfully
detected (AGB) OH/IR stars in the LMC. However, stronger than average
interference and time constraints resulted in rather noisy spectra (Fig.\ 7;
for comparison, spectra smoothed by a factor 11 are offset by a small amount).
Hints of emission features were not generally stable with time and are hence
deemed spurious until confirmed. IRAS\,04509$-$6922 was observed at a similar
noise level by Wood et al.\ (1992), who also failed to detect OH maser
emission in this source. The H~{\sc i} emission occurs at velocities in the
range 260 to 280 km s$^{-1}$ for IRAS\,04509$-$6922 and IRAS\,04516$-$6902,
and 220 to 270 km s$^{-1}$ for IRAS\,05294$-$7104.

%
% FIGURE 7
%
\begin{figure}
\centerline{\psfig{figure=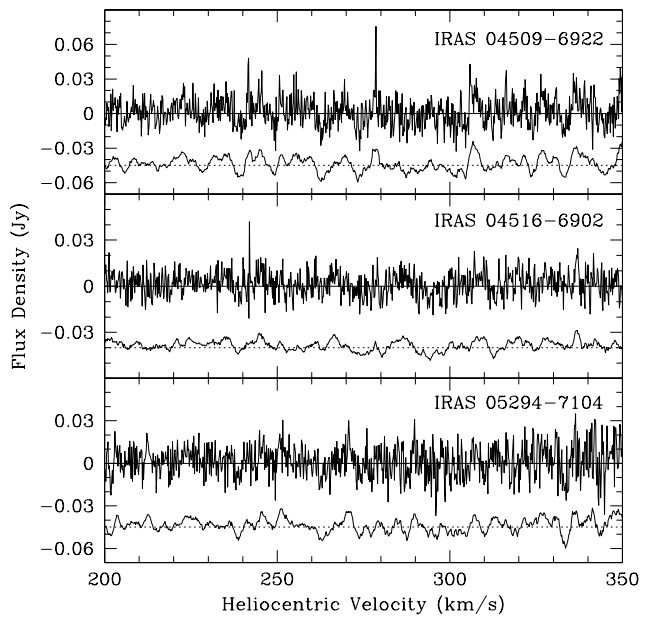,width=84mm}}
\caption[]{OH maser non-detections in three LMC targets.}
\end{figure}

Because there was no clear indication of maser emission in either of the SMC
targets after the first integrations, further effort was concentrated on the
LMC sources. Hence no circumstellar OH masers were found in the SMC, although
future observations might confirm potential emission between 130 and 150 km
s$^{-1}$ in IRAS\,00483$-$7347 (Fig.\ 8; the offset spectra are smoothed by a
factor 11). The H~{\sc i} emission covers a much wider range in velocities
than in the LMC, from about 100 to 190 km s$^{-1}$ in the direction of
IRAS\,00483$-$7347 and 120 to 200 km s$^{-1}$ in the direction of
IRAS\,00591$-$7307.

%
% FIGURE 8
%
\begin{figure}
\centerline{\psfig{figure=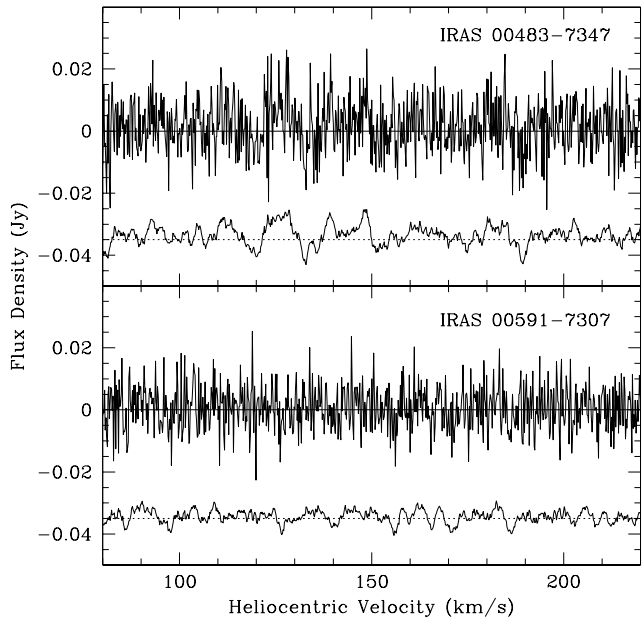,width=84mm}}
\caption[]{OH maser non-detections in two SMC targets.}
\end{figure}

%=========================================================================== 5
\section{Discussion}

%========================================================================= 5.1
\subsection{Galactic comparison samples}

\subsubsection{Galactic centre}

The only reasonably homogeneous comparison sample of OH/IR stars at a known
distance, the OH/IR stars in the galactic centre region of the Milky Way
galaxy provide the best comparison for the enlarged sample of OH/IR stars in
the LMC. The sample is described in Wood, Habing \& McGregor (1998) who
identified IR counterparts, and measured their bolometric luminosities and
pulsation periods, of OH maser sources from the galactic centre surveys by
Lindqvist et al.\ (1992) and Sjouwerman et al.\ (1998). We adopt a distance to
the galactic centre of $d_{\rm GC}=8$ kpc.

Blommaert et al.\ (1998), on the basis of similar data, argue for the presence
of a metal-rich ($\sim3$ times solar) population of OH/IR stars with $v_{\rm
exp}\gsim18$ km s$^{-1}$, and a somewhat less metal-rich and possibly older
population of OH/IR stars with $v_{\rm exp}\lsim18$ km s$^{-1}$, a conclusion
that was supported by Wood et al.\ (1998). We note, however, that the galactic
centre objects are significantly fainter and therefore probably have
significantly lower initial masses than the LMC objects.

\subsubsection{Galactic disk}

Galactic disk samples represent a mixture of stellar properties in between
those of the Magellanic Cloud and galactic centre populations. Although this
may lead to less ambiguous conclusions, it may also bridge the gap in
parameterspace between the fairly massive metal-poor OH/IR stars in the LMC
and the less-massive metal-rich OH/IR stars in the galactic centre.

We here compare with two galactic disk samples: (i) te Lintel Hekkert et al.\
(1991), and (ii) Groenewegen et al.\ (1998, their Appendix B). The Groenewegen
et al.\ sample also contains information about the pulsation periods, but the
te Lintel Hekkert et al.\ sample contains more OH masers. There is real
overlap between sources in these two samples, and both samples can be
considered equally representative of the galactic disk population of OH/IR
stars, with a natural bias towards sources that lie within the solar circle.

%========================================================================= 5.2
\subsection{Blue-red asymmetry}

Within our sample of LMC OH/IR stars, where both blue and red peaks are
detected, the blue peak is usually stronger than the red peak --- exceptions
to this are IRAS\,05329$-$6708 from Wood et al.\ (1992) which has a red peak
that is stronger than the blue, and IRAS\,05558$-$7000 where both peaks are of
equal strength. This was also noted in van Loon et al.\ (2001b), who proposed
that the blue peak might include a contribution from amplified stellar light
in addition to the amplified circumstellar radiation (Fig.\ 9). Indeed, VLBI
observations of nearby, galactic OH and H$_2$O masers indicate the possibility
for maser amplification of radiation originating from the stellar radiosphere
(Vlemmings, van Langevelde \& Diamond 2002; Vlemmings et al.\ 2003).

This predominant blue-asymmetry of the OH maser peaks in the LMC objects is in
stark contrast to the rather even distribution of blue/red peak flux density
ratios found in the galactic centre OH/IR stars (Fig.\ 10). There may be a
dependence on the bolometric luminosity, with the most blue-asymmetric profile
found in the by far most luminous object (IRAS\,04553$-$6825). Two effects may
be at work here: (i) the LMC objects are more luminous than the galactic
centre objects, which means that any contribution from amplified stellar light
is expected to be greater in the LMC objects, and (ii) the lower metallicity
--- and hence lower dust-to-gas ratio --- in the LMC objects may cause the
contribution from amplified circumstellar radiation to be smaller than in the
galactic centre objects.

%
% FIGURE 9
%
\begin{figure}
\centerline{\psfig{figure=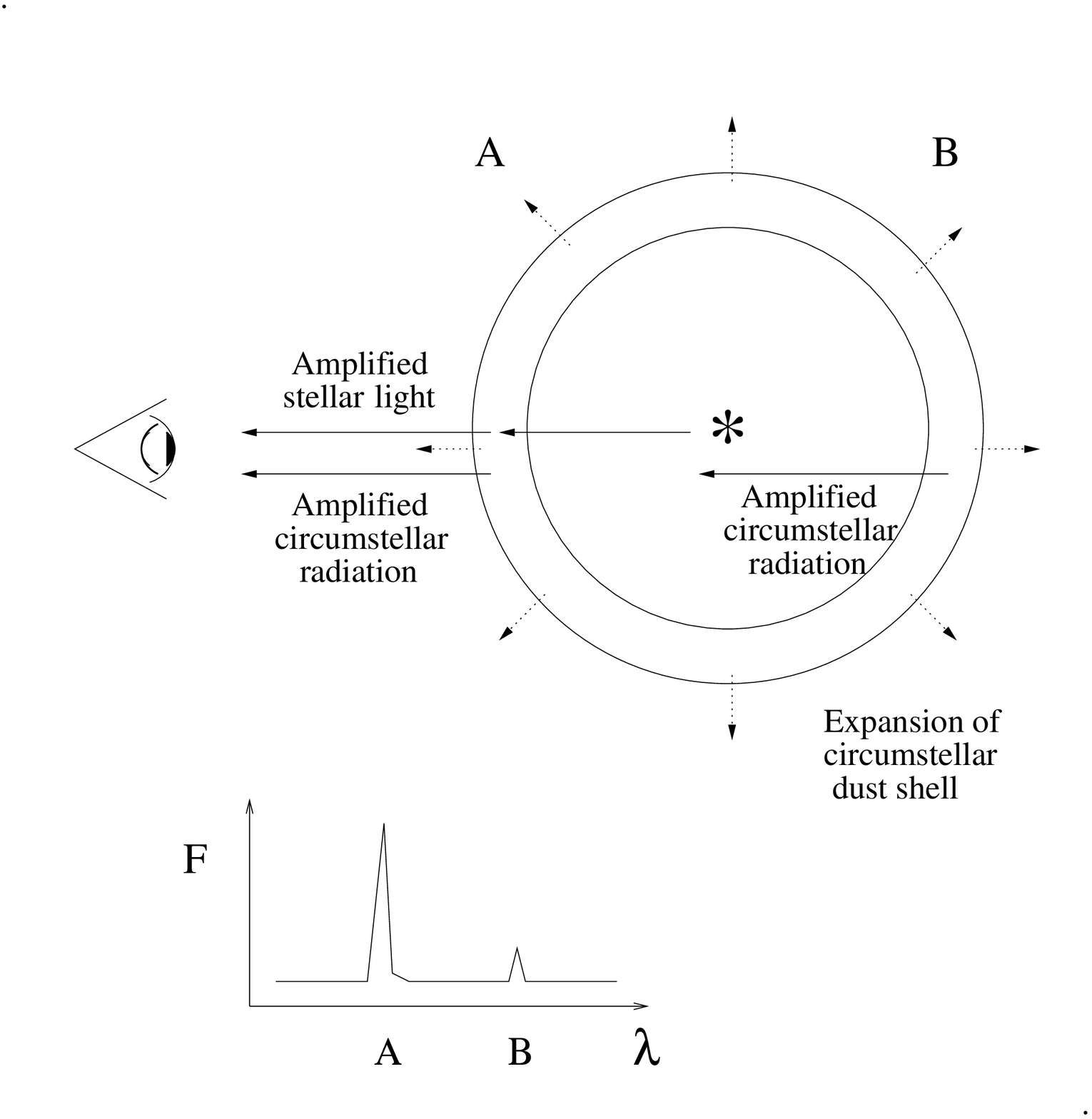,width=84mm}}
\caption[]{Explanation for the blue asymmetry of the OH maser profile as due
to the contribution from amplified stellar radiation. Shown are the maser
components which contribute to the observed spectrum.}
\end{figure}

%
% FIGURE 10
%
\begin{figure}
\centerline{\psfig{figure=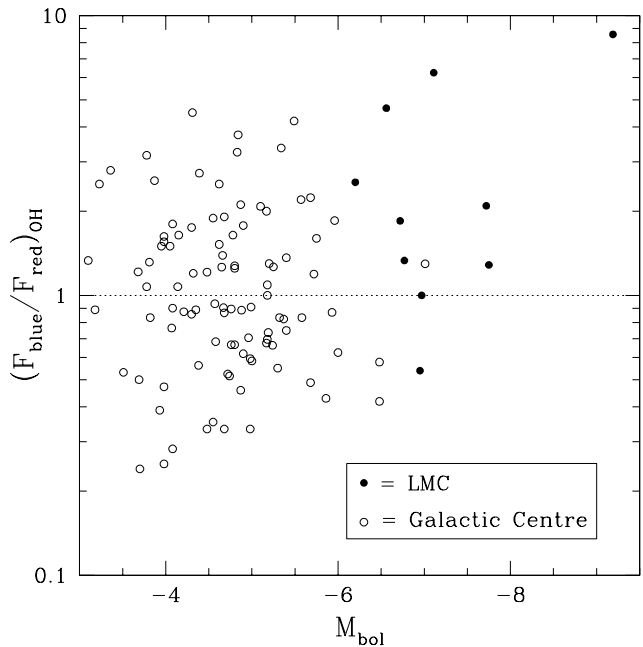,width=84mm}}
\caption[]{The ratio of the blue and red OH maser peak flux densities, versus
bolometric luminosity, for OH/IR stars in the LMC (dots) and galactic centre
(circles).}
\end{figure}

Competition between radiation which propagates outwards through the
circumstellar envelope and that which propagates inwards may enhance the
blue-asymmetry in the observed OH maser profile if the outwards propagating
maser utilises most of the population-inverted OH molecules, leaving little
for the inwards propagating maser. This would further diminish the red maser
component which results from inwards travelling radiation through the receding
part of the circumstellar envelope. Another consequence of this mechanism
would be that the amplification of the red maser component becomes less
radially beamed. This increased isotropy further diminishes the strength of
the observed maser radiation and might lead to underestimation of the wind
speed as inhomogeneities in the outflow yield similarly faint maser spots at a
variety of projected radial velocities.

Our findings are corroborated by data on galactic disk OH masers (Fig.\ 11).
The most red-asymmetric OH maser profiles have a ratio of blue and red OH
maser peak flux density of no less than 1:6. However, there are several
examples of OH maser profiles with this ratio in excess of 6:1. This suggests
that extremely blue-asymmetric profiles are not merely the tail of an
otherwise symmetric distribution of blue over red peak flux density ratios. On
average the wind speeds of these extremely blue-asymmetric OH masers are
indistinguishable from the bulk of the OH masers, although there is a
conspicuous lack of such sources with wind speeds below 10 km s$^{-1}$ (Fig.\
11, top panel). They are, however, significantly brighter (on average) OH
masers than the bulk (Fig.\ 11, centre panel). Their exclusively low galactic
latitudes, $|b_{\rm II}|<5^\circ$ (Fig.\ 11, bottom panel), strongly suggest
that these extreme OH masers are also more distant and hence more luminous in
an absolute sense.

OH/IR stars with the fastest winds, $v_{\rm exp}>22$ km s$^{-1}$ (Fig.\ 11,
top panel), are predominantly blue-asymmetric, in particular the two objects
with $v_{\rm exp}>30$ km s$^{-1}$. Such fast winds are most likely due to the
high luminosity of the central star. Indeed, some of the best-studied red
supergiants, NML\,Cyg, VY\,CMa, VX\,Sgr, S\,Per and IRC+10420 have fast
outflows, $v_{\rm exp}\sim20$ to 30 km s$^{-1}$, and they all have
blue-asymmetric profiles of 1612 MHz OH maser emission (Cohen et al.\ 1987;
Richards et al.\ 1999). A similar situation may occur in post-AGB objects
where a central source of radio emission in the form of an ionized wind may
dominate the maser amplification over the fainting, detached circumstellar
dust envelope itself (Zijlstra et al.\ 2001).

%
% FIGURE 11
%
\begin{figure}
\centerline{\psfig{figure=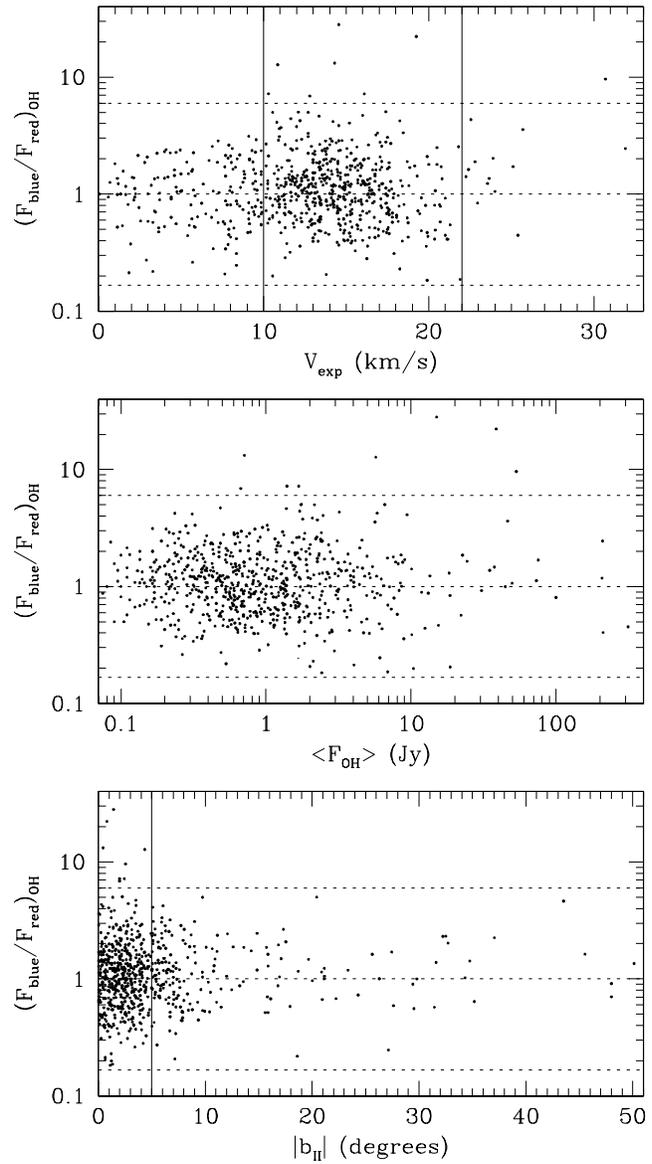,width=84mm}}
\caption[]{The ratio of the blue and red OH maser peak flux densities, versus
wind speed (top), mean OH maser peak flux (centre) and galactic latitude
(bottom), for OH/IR stars in the galactic disk (te Lintel Hekkert et al.\
1991). The dotted horizontal lines are for symmetric OH maser profiles and for
ratios of the blue and red OH maser peak flux densities of 6:1 and 1:6,
respectively. See text for an explanation of the vertical lines in the top and
bottom panels.}
\end{figure}

%========================================================================= 5.3
\subsection{OH maser strength}

In a lower metallicity environment the interstellar radiation field is
stronger and penetrates deeper into the circumstellar envelope (Lequeux et
al.\ 1994), which may lead to more dissociation of H$_2$O into OH. However, at
lower metallicity the abundance of H$_2$O in the circumstellar envelope may be
lower, resulting in a lower OH abundance. Do these effects cancel?

When plotted against bolometric magnitude (Fig.\ 12) the OH maser peak flux
densities of the LMC sample seem to follow from a direct scaling of the
galactic centre sample. This might be reconciled with similar OH abundances in
both populations of objects, where the difference in OH maser strength is only
due to a difference in the amount of radiation supplied by the embedded star
(either directly or via re-emission by the circumstellar dust). This may seem
remarkable given the fact that the abundance of the OH molecule also depends
on the mass-loss rate, which is shown to depend on luminosity (van Loon et
al.\ 1999) --- but perhaps not on metallicity (van Loon 2000). However, if the
masers are saturated then the rate at which 1612 MHz OH photons are emitted is
expected to be proportional to the pumping photon flux and hence (see below)
the infrared flux, which is the bolometric luminosity for an optically thick
envelope. In this case the metallicity does not matter.

%
% FIGURE 12
%
\begin{figure}
\centerline{\psfig{figure=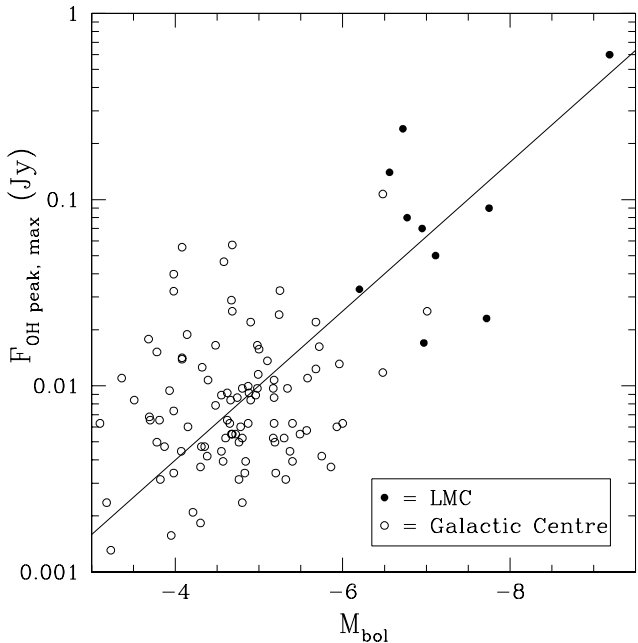,width=84mm}}
\caption[]{OH peak flux density versus bolometric luminosity, for OH/IR stars
in the LMC (dots) and galactic centre (circles). The straight line corresponds
to $F_{\rm OH peak, max}{\propto}L$.}
\end{figure}

The OH maser pump efficiency may be expressed as the ratio of the OH maser
peak flux density over the flux density of the 35 $\mu$m OH line. This line is
believed to be responsible (in part) for establishing the population inversion
in the OH molecule which allows for the maser mechanism to operate (Elitzur et
al.\ 1976). In the absence of direct measurements for the 35 $\mu$m flux
density the {\it IRAS} 25 $\mu$m flux density is taken instead. For OH/IR
stars in the galactic disk this leads to efficiencies in the region of
${\langle}F\rangle/F_{25}\sim$1\% to $\sim$20\%. With efficiencies of
${\langle}F\rangle/F_{25}\sim6\pm4$\% the LMC sources (Fig.\ 13, circles) are
indistinguishable from galactic disk objects (Fig.\ 13, dots). There is no
indication for the efficiency to be higher (or lower) in sources of
blue-asymmetric OH maser profiles. This suggests that the 1612 MHz OH masers
in both the galactic and LMC objects are indeed saturated.

A significant contribution to the scatter observed in Figs.\ 10 \& 12 is
due to variability, both in bolometric luminosity and OH maser intensity.
Whitelock et al.\ (2003) and Wood et al.\ (1992, 1998) derived time-averaged
bolometric magnitudes on the basis of near-IR photometric monitoring data, but
for the most extremely dust-enshrouded sources a significant fraction of the
total luminosity emerges at mid-IR wavelengths at which no such monitoring
data is available. The values for $M_{\rm bol}$ may thus be uncertain by up to
a few tenths of a magnitude, but this is accurate enough for the purpose of
the analyses performed here. The OH maser spectra were generally obtained only
once per object. The peak intensity of OH masers is known to vary by typically
a factor two (Herman \& Habing 1985; van Langevelde et al.\ 1993) and, whilst
contributing to the observed scatter, will not inhibit the detection of trends
or a statistical comparision between the magellanic and galactic samples.

%
% FIGURE 13
%
\begin{figure}
\centerline{\psfig{figure=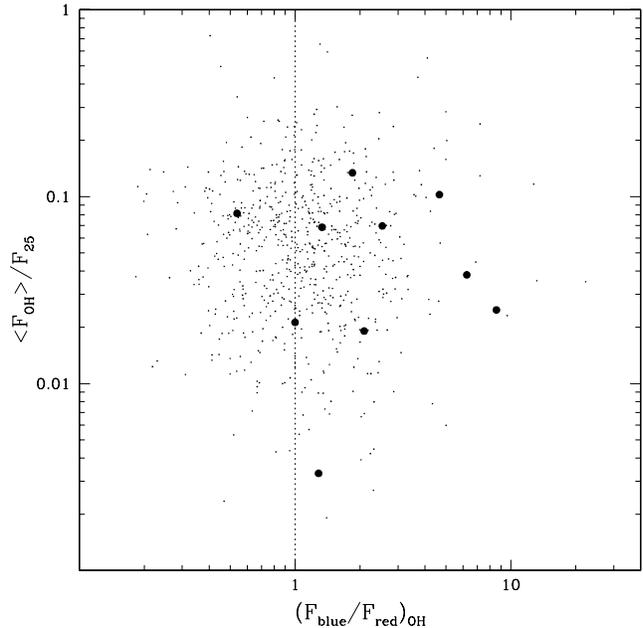,width=84mm}}
\caption[]{OH maser pump efficiency expressed as the ratio of mean OH maser
peak flux density over {\it IRAS} 25 $\mu$m flux density, versus the ratio of
the blue and red OH maser peak flux densities, for OH/IR stars in the LMC
(circles) and galactic disk (dots; te Lintel Hekkert et al.\ 1991). The dotted
vertical line is for symmetric OH maser profiles.}
\end{figure}

%========================================================================= 5.4
\subsection{Mass-loss rates from OH maser intensities}

If the OH maser is saturated then the peak flux density depends on the number
of population-inverted OH molecules, and may thus be a measure of the
mass-loss rate. Baud \& Habing (1983) proposed a prescription for this, based
on a calibration against OH/IR stars in the galactic disk. We use their
prescription to predict the mass-loss rates for OH/IR stars in the LMC, in a
form presented by van der Veen \& Rugers (1989):
\begin{equation}
\dot{M} = 1.8\times10^{-7} \sqrt{F_{\rm OH}} \ v_{\rm exp} D,
\end{equation}
where $\dot{M}$ is the mass-loss rate in M$_\odot$ yr$^{-1}$, $F_{\rm OH}$ is
the OH 1612 MHz maser peak intensity in Jy, $v_{\rm exp}$ is the wind speed in
km s$^{-1}$ and $D$ is the distance in kpc (Fig.\ 14, dotted line). These
predictions can be compared with the mass-loss rates as obtained from modeling
of the spectral energy distributions by van Loon et al (1999) (Fig.\ 14,
points). The predicted mass-loss rates are clearly too low, but they can be
brought in line with the measured mass-loss rates by reducing the OH abundance
by a factor 5 (Fig.\ 14, solid line) --- in Eq.\ (2) the OH abundance was
assumed to be $f_{\rm OH}=1.6\times10^{-4}$ (see Goldreich \& Scoville 1976).
This would suggest that the dominant factor in determining the OH abundance in
the superwind of OH/IR stars is the oxygen abundance (i.e.\ metallicity),
rather than the strength of the interstellar radiation field to dissociate
H$_2$O.

Zijlstra et al.\ (1996), based on a comparison with mass-loss rates as derived
from the IR dust emission, proposed an alternative prescription for the {\it
dust} mass-loss rate:
\begin{equation}
\dot{M}_{\rm dust} = 1.6\times10^{-10} F_{\rm OH} \ v_{\rm exp} D^2.
\end{equation}
The rationale behind this recipe is that if the OH maser is saturated then the
maser flux depends on the IR flux which pumps the population inversion, and
the OH maser is thus a measure of the dust mass-loss rate. Its predictions for
the LMC (Fig.\ 15, dotted line) fall below the measured values (Fig.\ 15,
points) by a factor 2, which may be accounted for by systematic uncertainties
in the modeling of the IR emission --- both for the LMC objects as well as the
galactic data that was used to calibrate Eq.\ (3).

The LMC data obey a relation of the type $\dot{M} \propto F_{\rm OH}$ (Eq.\
(3), Fig.\ 15) better than one of the type $\dot{M} \propto \sqrt{F_{\rm OH}}$
(Eq.\ (2), Fig.\ 14). This is unfortunate as it deprives us from a means to
measure the gas mass-loss rate if the OH maser emission yields the dust
mass-loss rate instead.

%
% FIGURE 14
%
\begin{figure}
\centerline{\psfig{figure=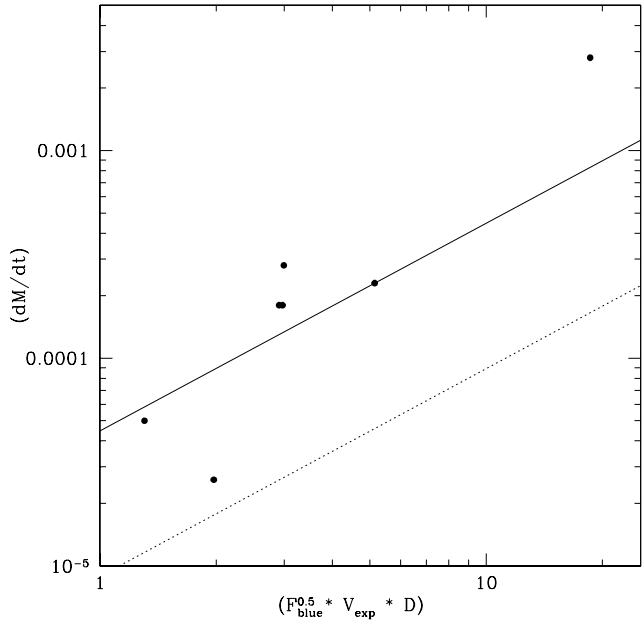,width=84mm}}
\caption[]{Predicted mass-loss rates following the Baud \& Habing (1983)
recipe (Eq.\ (2), dotted line) and LMC data with mass-loss rates from van Loon
et al.\ (1999) (points). The predictions can be scaled up to globally fit the
data (but not the slope) by reducing the OH abundance by a factor 5 (solid
line).}
\end{figure}

%
% FIGURE 15
%
\begin{figure}
\centerline{\psfig{figure=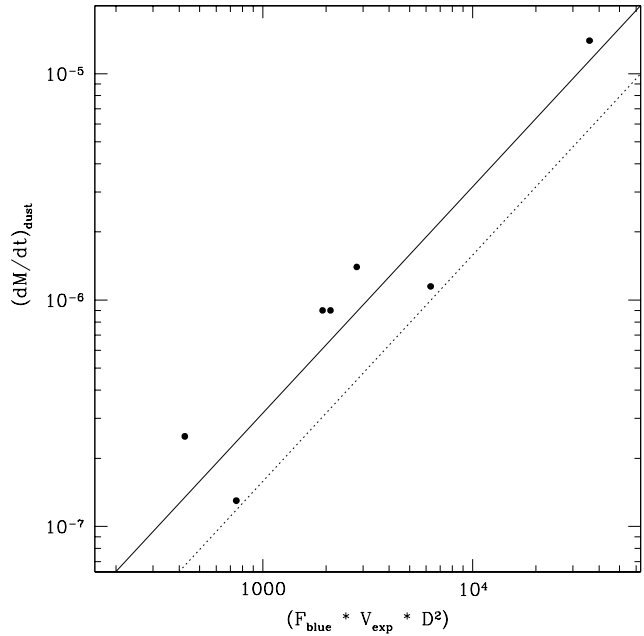,width=84mm}}
\caption[]{Predicted dust mass-loss rates following the Zijlstra et al.\
(1996) recipe (Eq.\ (3), dotted line) and LMC data with mass-loss rates from
van Loon et al.\ (1999) (points). The predictions can be scaled up by a factor
2 to fit the data (solid line).}
\end{figure}

%========================================================================= 5.5
\subsection{The wind speed}

We find a spread in wind speed comparable to that found in the galactic centre
(Figs.\ 16 \& 17). In general, higher velocities are found associated with the
most luminous stars (Fig.\ 14). It is clear, however, that at the same
luminosity, stars in the LMC have a lower wind speed. We can reproduce these
trends with Eq.\ (1), which is consistent for dust-to-gas ratioes $\Psi_{\rm
GC}/\Psi_{\rm LMC}\simeq6.5$ (dotted sequences in Fig.\ 16). This is very
similar to the ratio of the metallicities of the metal-rich galactic centre
sample and metal-poor LMC sample.

Most of the OH/IR stars in the galactic centre sample have shorter pulsation
periods than those in the LMC (Fig.\ 17). The difference in speed is smaller
between stars of comparable period than that shown in Fig.\ 16. In particular,
the OH/IR stars with the longest pulsation periods do not generally have the
fastest winds. This may be related to the changes in the stellar structure as
a result of the mass lost. It is important to bear in mind, though, that the
samples differ on at least one more crucial point: their progenitor masses.
Indeed, on average, at similar pulsation periods the wind speeds of the
Groenewegen et al.\ galactic disk sample are intermediate between the
Magellanic Cloud and galactic centre stars.

%
% FIGURE 16
%
\begin{figure}
\centerline{\psfig{figure=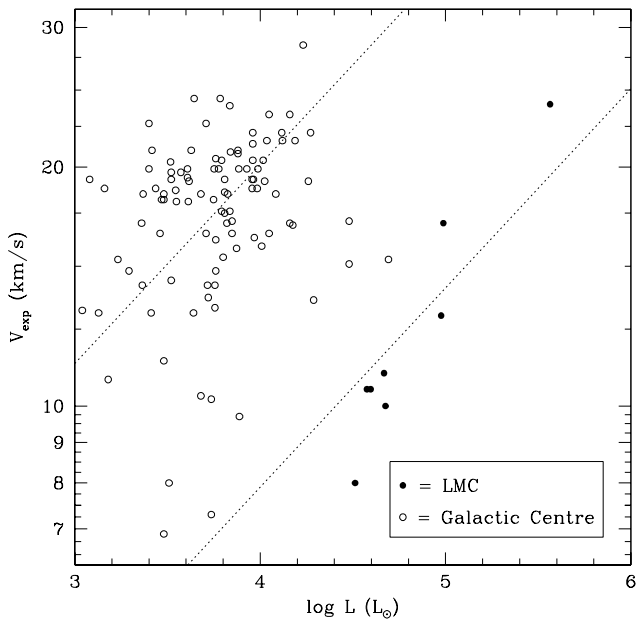,width=84mm}}
\caption[]{Wind speed versus bolometric luminosity, for OH/IR stars in the LMC
(dots) and galactic centre (circles). The dotted sequences follow Eq.\ (1) for
$\Psi_{\rm GC}/\Psi_{\rm LMC}\simeq6.5$.}
\end{figure}

%
% FIGURE 17
%
\begin{figure}
\centerline{\psfig{figure=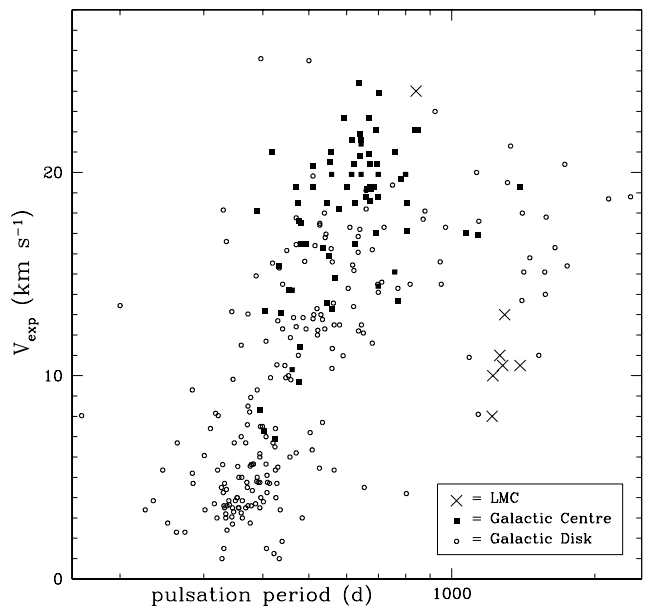,width=84mm}}
\caption[]{Wind speed versus period (after van Loon 2000), for OH/IR stars in
the LMC (crosses), galactic centre (filled squares) and galactic disk
(circles; Groenewegen et al.\ 1998).}
\end{figure}

%========================================================================= 5.6
\subsection{Comparison of wind speed and escape velocity}

We now compare the wind speed with the escape velocity from the "surface" of
the star, from the dust formation zone and from further out in the wind where
the OH masers originate. The stellar radius, $R_\star$, is determined from the
effective temperature, $T_{\rm eff}$ (estimated from the optical spectrum),
and the luminosity, $L$ (from $M_{\rm bol}$). The radius of the dust formation
zone, $R_{\rm dust}$, was obtained from detailed modelling of the spectral
energy distribution by van Loon et al.\ (1999). For the fully developed wind
we adopt $R_{\rm wind}=5{\times}R_{\rm dust}$.

%
% TABLE 4
%
\begin{table*}
\caption[]{Estimates for stellar mass and radius (see text), and the escape
velocities at the stellar surface, at the dust formation zone (van Loon et
al.\ 1999) and at a distance $5\times$ as far. The wind speed is compared with
these local escape velocities.}
\begin{tabular}{lccccccccccc}
\hline\hline
IRAS                      &
$M$                       &
$L$                       &
$T_{\rm eff}$             &
$R_\star$                 &
$R_{\rm dust}$            &
\multicolumn{3}{c}{$v_{\rm esc}$ (km s$^{-1}$)} &
\multicolumn{3}{c}{$v_{\rm exp}/v_{\rm esc}$}   \\
\multicolumn{6}{c}{\rule[1mm]{0mm}{0mm}}      &
\multicolumn{3}{c}{\rule[1mm]{36mm}{0.2mm}}   &
\multicolumn{3}{c}{\rule[1mm]{36mm}{0.2mm}}   \\
name                      &
(M$_\odot$)               &
($10^3$ L$_\odot$)        &
(K)                       &
($10^3$ R$_\odot$)        &
($10^3$ R$_\odot$)        &
[$R_\star$]               &
[$R_{\rm dust}$]          &
[$5{\times}R_{\rm dust}$] &
[$R_\star$]               &
[$R_{\rm dust}$]          &
[$5{\times}R_{\rm dust}$] \\
\hline
04407$-$7000           &
6                      &
55                     &
3008                   &
0.87                   &
7.7                    &
51.4                   &
17.2                   &
7.7                    &
                       &
                       &
                       \\
04498$-$6842           &
8                      &
97                     &
2500                   &
1.66                   &
\llap{1}4.8            &
42.8                   &
14.4                   &
6.4                    &
0.30                   &
0.91                   &
2.02                   \\
04509$-$6922           &
7                      &
65                     &
2500                   &
1.36                   &
8.1                    &
44.3                   &
18.1                   &
8.1                    &
                       &
                       &
                       \\
04516$-$6902           &
6                      &
55                     &
2667                   &
1.10                   &
7.7                    &
45.5                   &
17.2                   &
7.7                    &
                       &
                       &
                       \\
04545$-$7000           &
4                      &
33                     &
2890\rlap{$^1$}        &
0.73                   &
8.8                    &
45.7                   &
13.2                   &
5.9                    &
0.18                   &
0.61                   &
1.36                   \\
04553$-$6825           &
\llap{3}0              &
\llap{3}77             &
3008                   &
2.26                   &
\llap{3}6.2            &
71.2                   &
17.8                   &
8.0                    &
0.34                   &
1.35                   &
3.02                   \\
05003$-$6712           &
4                      &
24                     &
2667                   &
0.73                   &
6.9                    &
45.9                   &
14.9                   &
6.7                    &
                       &
                       &
                       \\
05280$-$6910           &
\llap{1}5              &
\llap{1}00             &
2890\rlap{$^1$}        &
1.26                   &
\llap{1}7.7\rlap{$^2$} &
67.3                   &
18.0                   &
8.0                    &
0.25                   &
0.95                   &
2.11                   \\
05294$-$7104           &
5                      &
41                     &
2890                   &
0.81                   &
8.0                    &
48.5                   &
15.4                   &
6.9                    &
                       &
                       &
                       \\
05298$-$6957           &
4                      &
39                     &
2890\rlap{$^1$}        &
0.79                   &
9.8                    &
44.1                   &
12.7                   &
5.6                    &
0.24                   &
0.84                   &
1.88                   \\
05329$-$6708           &
5                      &
48                     &
2890\rlap{$^1$}        &
0.87                   &
\llap{1}0.5            &
46.7                   &
13.5                   &
6.0                    &
0.24                   &
0.82                   &
1.82                   \\
05402$-$6956           &
5                      &
41                     &
2890\rlap{$^1$}        &
0.80                   &
9.4                    &
48.7                   &
14.2                   &
6.4                    &
0.22                   &
0.74                   &
1.65                   \\
05558$-$7000           &
5                      &
49                     &
2890\rlap{$^1$}        &
0.88                   &
8.9                    &
46.5                   &
14.6                   &
6.5                    &
0.22                   &
0.69                   &
1.53                   \\
\hline
\multicolumn{5}{l}{\it Galactic centre comparison sample} \\
Average                &
1                      &
5\rlap{.2}             &
2890\rlap{$^1$}        &
0.29                   &
                       &
36.3                   &
                       &
                       &
0.46                   &
                       &
                       \\
\hline
\end{tabular}\\
\flushleft{Notes: $^1$ adopted spectral type M8; $^2$ adopted $R_{\rm
dust}=14{\times}R_\star$.}
\end{table*}

The masses of these objects are uncertain. However, we can assume masses of 4
to 8 M$_\odot$ as lower-mass objects would have become carbon stars and 8
M$_\odot$ corresponds to the upper limit for AGB evolution. IRAS\,04553$-$6825
is a supergiant and will therefore be more massive ($\sim$15 to 50 M$_\odot$).
IRAS\,05280$-$6910 is in the $\sim10^7$-year old cluster NGC\,1984 and will
have had a mass of $\sim$15 M$_\odot$ (Wood et al.\ 1992). IRAS\,05298$-$6957
is also in a (small) cluster, HS\,327, and had a mass of $\sim$4 M$_\odot$
(van Loon et al.\ 2001a). The masses for the other objects have been estimated
on the basis of their luminosity. The adopted and calculated values for the
masses, radii and escape velocities are listed in Table 4.

It is interesting to note that the escape velocity at the stellar surface is
with $v_{{\rm esc},\star}\simeq43$ to 48 km s$^{-1}$ very similar for all
tip-AGB stars. The escape velocities for the supergiants IRAS\,04553$-$6825
($v_{{\rm esc},\star}=71$ km s$^{-1}$) and IRAS\,05280$-$6910 ($v_{{\rm
esc},\star}\sim67$ km s$^{-1}$) are also very similar but significantly faster
than those of the AGB stars. The galactic centre stars are almost certainly of
significantly lower mass, and although their optical spectral types are not
known there is no reason to believe that they should be on average warmer than
M8. It is therefore very probable that the escape velocity at their stellar
surface (here estimated as 36 km s$^{-1}$) is indeed lower than for the
magellanic AGB stars of our sample.

At the speed as measured from the OH maser, the material would not have
escaped from the stellar surface nor from the dust formation radius --- except
for the extremely luminous supergiant IRAS\,04553$-$6825 --- without further
driving (Fig.\ 18). Netzer \& Knapp (1987) show that the location of a maser
of molecules produced by photodissociation (e.g.\ OH) is $R_{\rm OH}>2R_{\rm
dust}$. At this distance, the wind speed does exceed the local escape velocity
for all OH/IR stars in the LMC (Fig.\ 18, dotted lines). Remarkably, this
critical point seems to be very nearly identical for all OH/IR stars, at an
``escape radius'' $R_{\rm esc}\simeq20,000{\times}R_\odot$, despite their
differences in dust formation radii and hence acceleration timescales before
reaching the escape radius. This scale may thus be fundamental to driving a
steady outflow, and might result from the self-regulatory nature of
dust-driven winds (Struck et al.\ 2004).

The galactic centre stars have with an average wind speed of 16.7 km s$^{-1}$
a relatively fast wind, which in combination with their low escape velocities
results in a smaller escape radius of $R_{\rm esc}\sim$1,400 R$_\odot$. As the
escape radius is similar for magellanic objects of vastly different masses
(and luminosities), the most obvious stellar parameter that could be
responsible for the differences between the escape radii in the LMC and the
galactic centre is the metallicity.

We note that the current masses of the OH/IR stars are less than their
progenitor masses due to the intense mass loss that made them OH/IR objects in
the first place. Although the 12 and 25 $\mu$m emission which led {\it IRAS}
to detect the LMC sources is very bright, it originates from warm dust which
was lost recently. Even the OH maser emission arises predominantly from
material which was ejected $t\sim10^3$ yr ago (see van der Veen \& Rugers
1989). At a steady mass-loss rate of $\dot{M}\sim10^{-5}$ M$_\odot$ yr$^{-1}$,
this amounts to an ejected mass only of the order ${\Delta}M\sim0.01$
M$_\odot$. If, however, the OH/IR phase of evolution is where these stars lose
most of their mass then statistically we can expect a (relatively massive)
OH/IR star to have lost on average $\sim40$\% of their progenitor mass. This
would reduce our estimates for their escape velocities by a factor
$\sim\sqrt{0.6}$ and increase the ratios of $v_{\rm exp}/v_{\rm esc}$ by the
same factor $\sim1.3$ at all radii. Thus the stellar wind may in fact already
have reached the escape velocity very near the dust formation zone (see Fig.\
18). The reduced masses would lead to a reduction in the value for the escape
radius for LMC stars to $R_{\rm esc}\simeq10,000{\times}R_\odot$. However,
statistically an identical reduction would apply to all LMC stars, which would
therefore still share a surprisingly common value for this lengthscale and one
that is still significantly longer than for galactic centre stars.

%
% FIGURE 18
%
\begin{figure}
\centerline{\psfig{figure=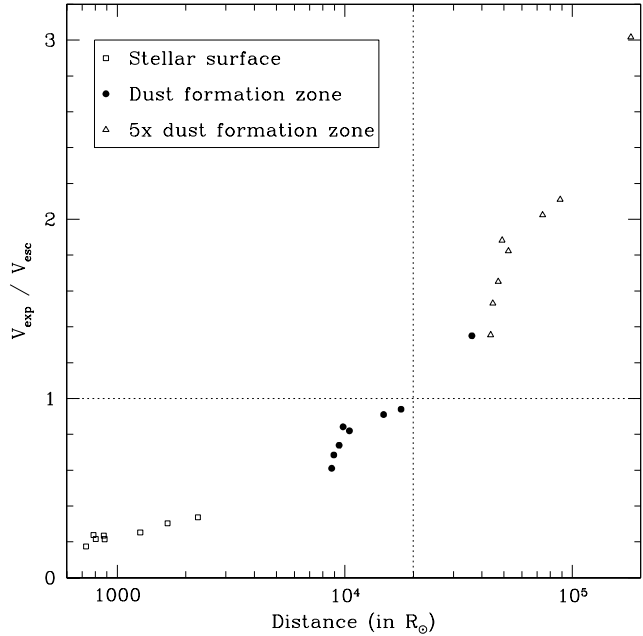,width=84mm}}
\caption[]{Ratio of measured wind speed and estimated local escape velocity,
versus the distance from the central star at which the escape velocity was
evaluated.}
\end{figure}

%=========================================================================== 6
\section{Conclusions}

We have presented the results of a new survey for OH maser emission at 1612
MHz in the circumstellar environments of magellanic dust-enshrouded giants.

Of the eight LMC sources selected, five were found to have OH masers, with at
least one of the non-detections indicating that a detection is possible with
increased integration time. This suggests that the sample was well chosen, due
to the selection criteria employed: long pulsation period ($P\gsim800$ d),
large pulsation amplitude (${\Delta}K\gsim1.2$ mag), late oxygen-rich spectral
type ($\gsim$M8), red ($J-K\gsim2$ mag), and bolometrically and mid-IR bright
($M_{\rm bol}\lsim-6$ mag and $F_{\rm 25}\gsim0.3$ Jy, respectively). We can
thus be confident in our selection of future potential OH/IR stars. The
strength of the OH maser peak in the LMC detections can be readily scaled from
the strength of the OH maser peak in galactic centre sources on the basis of
the differences in bolometric luminosity alone, without the need for an
additional scaling due to the difference in metallicity. This provides some
comfort for future searches for OH masers in the SMC or other low-metallicity
environments.

The LMC sources show a pronounced asymmetry between the strength of the blue
and redshifted emission, much more so than OH/IR stars in the galactic centre.
We propose that this is due to a greater contribution of amplified radiation
originating from the star itself, both in an absolute sense (bolometrically
brighter stars) and because the lower metallicity of sources in the LMC leads
to a smaller contribution of (amplified) circumstellar emission compared with
OH/IR stars in the galactic centre.

Of the recipes for deriving mass-loss rates from the OH maser emission from
Baud \& Habing (1983) for the gas mass-loss rate and from Zijlstra et al.\
(1996) for the dust mass-loss rate the LMC data follow the Zijlstra et al.\
predictions better than the Baud \& Habing predictions. This supports the
interpretation of Zijlstra et al.\ that the OH maser strength is a measure of
the IR pumping flux and thus of the dust mass-loss rate.

We find that at the same luminosity or pulsation period, the difference in
wind speed, $v_{\rm exp}$, can be fully accounted for by metallicity effects.
This agrees with the suggestion by Wood et al.\ (1998) that the stars with
higher $v_{\rm exp}$ have higher metallicity. In addition, we show that the
current observations are consistent with simple radiation-driven wind theory
in which the wind speed depends on luminosity, $L$, and metallicity, $z$, as
$v_{\rm exp}{\propto}z^{1/2}L^{1/4}$.

We compare the wind speed with the escape velocities, evaluated at the stellar
surface, in the dust formation zone and further out in the wind. We find that
the wind speed exceeds the local escape velocity at an ``escape radius'' of
about 1--2$\times10^4$ R$_\odot$ from the star (just beyond the dust
condensation radius) --- a value which is surprisingly similar for all OH/IR
stars in the LMC, AGB stars and red supergiants alike (but different for
galactic centre OH/IR stars).

\section*{Acknowledgments}

We thank the staff at Parkes Observatory for their support and hospitality,
and Mark Calabretta for correspondence regarding SPC. An anonymous referee is
thanked for her/his constructive and useful comments. JRM is supported by a
PPARC studentship, and MM is supported by a PPARC Rolling Grant.

\label{lastpage}

\end{document}